\documentclass[a4paper,11pt]{article}
\usepackage{jheppub}
\usepackage[T1]{fontenc}
\usepackage{lmodern}
\usepackage{amsmath}
\usepackage{amsfonts}
\usepackage{amssymb}
\usepackage{mathrsfs}
\usepackage{physics}
\usepackage{graphicx}
\usepackage{caption}
\usepackage{subcaption}
\usepackage{enumitem}
\usepackage{framed,float}
\usepackage{booktabs}
\usepackage{array}
\usepackage{soul}
\usepackage{multirow}
\usepackage{comment}
\usepackage{braket}
\usepackage{todonotes}
\usepackage{youngtab}
\usepackage{tikz}
\usepackage{color}
\usepackage{extarrows}
\usetikzlibrary{calc,arrows,decorations.markings}

\def\cG{{\mathbb{G}}}
\def\cF{{\mathbb{F}}}
\def\cq{{\mathfrak{q}}}
\def\cL{{\mathcal{L}}}

\newcommand{\xar}[1]{x^{(#1)}}
\newcommand{\Yar}[1]{Y^{(#1)}}

\newcommand{\dup}{\mathrm{d}}
\newcommand{\iup}{\hspace{1pt}\mathrm{i}\hspace{1pt}}
\newcommand{\fraco}[1]{\frac{1}{#1}}

\newcommand{\Int}{\int\limits}

\def\b{\beta}

\def\l{\lambda}
\def\m{\mu}

\def \be  {\begin{equation}}
\def \ee  {\end{equation}}
\def \ba {\begin{equation}\begin{aligned}}
\def \ea {\end{aligned}\end{equation}}
\def \bea  {\begin{eqnarray}}
\def \eea  {\end{eqnarray}}

\newcommand{\prodp}[1]{\prod_{(i,j)\in{#1}}}

\title{\boldmath Proof of 5D $A_n$ AGT conjecture at $\beta=1$}

\author[a]{Qian Shen,}
\author[a,b]{Zi-Hao Huang,}
\author[a,c]{Shao-Ping Hu,}
\author[a,d,1]{Qing-Jie Yuan,}
\author[a,e,f,1]{and Kilar Zhang\note{Corresponding author.}}

\affiliation[a]{Department of Physics and Institute for Quantum Science and Technology, Shanghai University, Shanghai 200444, China}
\affiliation[b]{School of Physics Science and Engineering, Tongji University, Shanghai 200092, China }
\affiliation[c]{Department of Applied
Physics, The Hong Kong Polytechnic University, Hong Kong 999077, China}
\affiliation[d]{Department of Physics and Astronomy, Uppsala University, Box 516, 75120 Uppsala, Sweden}
\affiliation[e]{Shanghai Key Lab for Astrophysics, Shanghai 200234, China}
\affiliation[f]{Shanghai Key Laboratory of High Temperature Superconductors, Shanghai 200444, China}

\emailAdd{shenqian@shu.edu.cn}
\emailAdd{huangzihao@shu.edu.cn}
\emailAdd{279579878@shu.edu.cn}
\emailAdd{wolfgangyuan@shu.edu.cn}
\emailAdd{kilar@shu.edu.cn}

\abstract{In this paper, we give a proof of 5D $A_n$ AGT conjecture at $\beta=1$, where the gauge theory side is one dimension higher than the original 4D case, and corresponds to the q-deformation of the 2D conformal field theory side. We define a q-deformed $A_n$ Selberg integral, which generalizes the $A_n$ Selberg integral and the q-deformed $A_1$ Selberg integral in the literature. A q-deformed $A_n$ Selberg average formula  with $n+1$ Schur polynomials is proposed and proved to complete the proof.}

\begin{document} 
\maketitle
\flushbottom
\section{Introduction}

The original Alday-Gaiotto-Tachikawa (AGT) conjecture \cite{aldayLiouvilleCorrelationFunctions2010,LeFloch:2020uop} shows the correspondence between 4D $\mathcal{N}=2$ gauge theory and 2D conformal field theory (CFT). In particular, the Nekrasov partition function is identified with the  correlation functions in CFT. Soon it is generalized to the so-called 5D AGT conjecture \cite{Awata:2009ur,aganagicGaugeLiouvilleTriality2013,aganagicA_nTriality2014}. In this case, the gauge theory 
 is lifted by one dimension higher, and corresponds to the q-deformation on the CFT side.
The process of generalization involves introducing a deformation parameter $q=e^{-\hbar R}$, where $R$ is the radius of the compact fifth dimension. The q-deformation essentially scales the parameters of the theory according to powers of $q$, which can be related to the extra dimension's compactification radius. Taking the $q\rightarrow1$ limit, namely $R\rightarrow0$, we can regain the 4D case.

Initially, a direct proof of the correspondence in 4D case of $A_1$ group at $\beta=1$\footnote{$\beta=-\epsilon_1/\epsilon_2$, where $\epsilon_1$ and $\epsilon_2$ depict the $\Omega$ background.} is given in \cite{Mironov:2010pi}. There are also recursive checks for general $\beta$ in \cite{albaCombinatorialExpansionConformal2011, Kanno:2013aha}. Recently, the proof of 4D $A_n$ case at $\beta=1$ is given in \cite{yuanProofA_AGT2023}, making use of the formula of Selberg integral for Schur polynomials conjectured in \cite{zhangSelbergIntegralSU2011} and proved in \cite{albionAFLTtypeSelbergIntegrals2021}. While for the 5D case, we need the q-Selberg integral for Schur polynomials (Macdonald polynomials in the case of general $\beta$). There are already some previous works along this direction in \cite{mironovProvingAGTConjecture2012,awataFivedimensionalAGTConjecture2010}, (and other methods like \cite{Fukuda:2019ywe,Zenkevich:2014lca,Taki:2014fva}), by illustrating the main idea of taking q-deformations on both sides of the 4D equations. The q-analogue of an arbitrary number $n$ is defined by
\be\label{nq}
[n]_q\equiv\frac{1-q^{n}}{1-q}.
\ee
Obviously, when $q$ tends to 1, the right side of the above equation reduces to $n$.
However, the q-deformation of an arbitrary equation is not trivial, since it does not necessarily preserve the product of the numbers after deformation. For example,
if we assume $n_1 n_2=m_1 m_2$ and perform q-deformation on both sides of the equation,  in general 
$[n_1]_q [n_2]_q\neq [m_1]_q [m_2]_q.$
Now that the q-deform procedure is a non-trivial generalization, in practice we still need to verify whether the equations remain hold afterwards. In this paper, we dwell on the details of this 5D version of proof at $\b=1$, and show that indeed the conjecture is a theorem. Mathematically, the 5D case involves q-W algebra, or equivalently Ding-Iohara-Miki algebra \cite{Ding:1996mq, miki2007q} and Yangian \cite{Ragoucy:1998bv,Brundan:2004ca,Matsuo:2023lky}, which play important roles in integrability.

The paper is organized as follows. In section 2, we introduce the partition function in 5D. In section 3, we give a brief introduction to the Selberg integral and find a q-deformed Selberg average formula. In section 4, we calculate the expression for the conformal blocks after q-deformation. In section 5,
we provide a direct proof at $\b=1$. Other necessary details are collected in the appendix.

\section{Instanton partition function in 5D}
We start with considering the instanton partition function of $\mathcal{N}=1$ 5D theory. To transform the 4D Nekrasov partition function introduced in \cite{yuanProofA_AGT2023} into a 5D one, all one needs is to perform a suitable q-deformation on all factors of the 4D partition function. 

For convenience of description, we define 
\begin{eqnarray}
\cF_Y(z,q)&=&\prod_{(i, j)\in Y} (1-q^{z+\beta(i-1)-(j-1)}),\\
\cG_{Y,W}(x,q)&=&\prod_{(i, j)\in Y}(1-q^{x+\beta(Y_j^{\prime}-i)+(W_i-j)+\beta}).
\end{eqnarray}
After performing the q-deformation defined by \eqref{nq}, the instanton part of the 5D Nekrasov partition function can be written in the form of instanton expansion \cite{awataFivedimensionalAGTRelation2010,awataRefinedBPSState2009}:
\be\label{5d partition}
\begin{split}
&Z_{\text{inst}}(\cq)=\sum_{\vec Y}
\cq^{|\vec Y|} N_{5d}^\mathrm{inst}( \vec a,  \mu),\\
N_{5d}^\mathrm{inst}&=\prod_{r,s=1}^{n+1}\frac{\cF_{Y^{(s)}} (\mu_r+a_s,q^{-1})\cF_{Y^{(s)}} (\mu_{n+1+r}+a_s,q)}{\cG_{Y^{(r)},Y^{(s)}}(a_r-a_s,q)\cG_{Y^{(s)},Y^{(r)}}(a_s-a_r+1-\b,q^{-1})},
\end{split}
\ee
where $\cq$ is the gauge coupling constant (not to be confused with the q-deformation), $\vec a_I$ is the adjoint scalar vacuum expectation value,  and $\m_i$ is the mass of (anti-)fundamental hypermultiplet.
$Y$ is the Young diagram, which is known as a graphical representation of a partition of a positive integer. A Young diagram consists of rows of boxes, aligned to the left, with each row containing a number of boxes that is non-increasing from top to bottom. The shape of the Young diagram corresponds to the partition it represents, with the total number of boxes $|Y|$ equaling the integer being partitioned. The arm length and leg length of a box \( (i,j) \) in the Young diagram are denoted by \( {\rm Arm}_Y(i,j) \) and \( {\rm Leg}_Y(i,j) \), which have the following forms
\be
{\rm Arm}_Y(i,j) = Y^{\prime}_j - i, \quad {\rm Leg}_Y(i,j) = Y_i - j,
\ee
where \( Y^{\prime} \) represents the transpose of the Young diagram, $Y_i$ is the height of the $i^\mathrm{th}$ column and $Y'_i$ is the length of the $i^\mathrm{th}$ row. $\vec{Y}=\{Y^{(1)},Y^{(2)}, \dots , Y^{(N)}\}$ represents a set of Young diagrams, which can be used to represent instantons. 

The 5D partition function has different equivalent forms through rescaling the expansion parameter $\cq$. Considering that
\be
\label{exchange-equal}
\cG_{Y,W}(x,q)\cG_{W,Y}(-x+1-\b,q^{-1})=\cG_{W,Y}(x,q)\cG_{Y,W}(-x+1-\b,q^{-1}),
\ee
our convention agrees with \cite{awataFivedimensionalAGTRelation2010,awataRefinedBPSState2009}.
Similarly, we also need to perform q-deformation on conformal blocks corresponding to the 5D theory, which will be discussed in subsequent sections.

%%%%%%%%%%%%
\section{q-deformed Selberg integral}\label{Selberg}
In this section, we discuss the Selberg integral for preparation of the representation of the q-deformed conformal blocks. We define a q-deformed $A_n$ Selberg integral and prove a new Selberg average formula.

The Selberg integral is a generalization of the Euler beta function, which is introduced by Atle Selberg
in the 1940s \cite{selberg1944bemerkninger,kadell1988proof}. It has profound implications in various fields and serves as a powerful tool for computing conformal blocks. In \cite{albaCombinatorialExpansionConformal2011}, a Selberg integral over a pair of Jack polynomials \cite{stanley1989some} is used to verify the relationship between conformal blocks in Liouville field theory and $\mathcal{N}=2$ supersymmetric gauge theory for $SU(2)$. Moreover, the Selberg integral has been utilized to derive explicit combinatorial expressions for conformal blocks. Different types and generalizations of Selberg integrals have also been explored, which include an $A_n$  analogue of the AFLT integral containing a product of Jack polynomials  in the integral, and an elliptic generalization of the AFLT integral where the symmetric polynomials have been replaced with elliptic interpolation functions \cite{albionAFLTtypeSelbergIntegrals2021,albionEllipticMathrm_n2023}. Besides, the q-Jackson integral is also a related integral form corresponding to the 5D situation. From 0 to $a$, the q-Jackson integrals is defined by
\be\label{jackson}
\int_0^a f(x) d_{q} x=(1-q) a \sum\limits_{k=0}^{\infty} q^k f( q^k a ).
\ee
In this paper, we define a more general q-deformed Selberg integral form to represent the q-deformed conformal blocks etc..

Based on the $A_n$ Selberg integral introduced in \cite{warnaarSelbergIntegralLie2009} and the q-deformed $A_1$ Selberg integral introduced in \cite{albionAFLTtypeSelbergIntegrals2021}, for a symmetric polynomial $\mathcal{O}$ composed of products of Macdonald polynomials $M_\l(x;q,t)$, we define a q-deformed $A_n$ Selberg integral
{\small
\be\label{q-integral}
\begin{split}
&I^{A_n}_{N_1,\dots,N_n}(\mathcal{O};u_1,\dots,u_n,v;\b)\\
=&\frac{1}{N_1!\cdots N_n!(2\pi\iup)^{N_1+\cdots+N_n}} \Int_{\mathbb{T}^{N_1+\cdots+N_n}}\mathcal{O}\big(x^{(1)},\dots,x^{(n)}\big)\prod_{i=1}^{N_n}\frac{(q^{a_n}/x^{(n)}_i,q^{1-a_n}\xar{n}_i;q)_{\infty}}
{(q^{v+1}/x^{(n)}_i,\xar{n}_i;q)_{\infty}}
\prod_{r=1}^{n-1}\prod_{i=1}^{N_r}(\xar{r}_i)^{a_r}\\
\times&\prod_{r=1}^n \prod_{1\leqslant i<j\leqslant N_r} \frac{(x^{(r)}_i/x^{(r)}_j,x^{(r)}_j/x^{(r)}_i;q)_{\infty}}
{(tx^{(r)}_i/x^{(r)}_j,tx^{(r)}_j/x^{(r)}_i;q)_{\infty}}\prod_{r=1}^{n-1}\prod_{i=1}^{N_r}\prod_{j=1}^{N_{r+1}}\frac{((qt)^{1/2}x^{(r+1)}_j/x^{(r)}_i;q)_{\infty}}
{((q/t)^{1/2}x^{(r+1)}_j/x^{(r)}_i;q)_{\infty}}
\frac{\dup x^{(1)}}{x^{(1)}}\cdots \frac{\dup x^{(n)}}{x^{(n)}},
\end{split}
\ee
}
where $\mathbb{T}$ is the positively oriented unit circle, $N_r$ is the length of $x^{(r)}_i$, $t=q^\b$, parameters $a_r=u_r+\b N_r+1-2\b$ for $r=1,\dots,n-1$ and $a_n=u_n+v+\b N_n+2-\b$. $u_r,v_r$ are $A_n$ Selberg integral parameters and $v_1=\cdots=v_{n-1}\equiv 0$, $v_{n}\equiv v$. Here we also use some notations of q-shifted factorials
\be\label{q-shift}
\begin{split}
(z_1,z_2;q)_{\infty}=(z_1;q)_{\infty}(z_2;q)_{\infty},
\quad
(z;q)_{\infty}=\prod_{k\geqslant0}(1-q^k z),\quad(z;q)_b=\frac{(z;q)_{\infty}}{(z q^b;q)_{\infty}}.    
\end{split}
\ee

To illustrate the rationality, we should check that the integral element may return to the original one when $q$ tends to 1. According to the limit \cite{askeyBasicHypergeometricExtensions1980,forresterImportanceSelbergIntegral2008}
\be\label{limit}
\lim_{{q \to 1}} \frac{(z;q)_\infty}{(zq^b;q)_\infty}  = (1-z)^b,
\ee
the integral element can be transformed as follows
\be\label{q-factor}
\frac{(q^{a_n}/x^{(n)}_i,q^{1-a_n}\xar{n}_i;q)_{\infty}}
{(q^{v+1}/x^{(n)}_i,\xar{n}_i;q)_{\infty}}\xrightarrow{q\to 1}(1-\fraco{x^{(n)}_i})^{v+1-a_n}(1-\xar{n}_i)^{a_n-1}.
\ee
Similarly,
\be
\frac{(x^{(r)}_i/x^{(r)}_j,x^{(r)}_j/x^{(r)}_i;q)_{\infty}}
{(tx^{(r)}_i/x^{(r)}_j,tx^{(r)}_j/x^{(r)}_i;q)_{\infty}}
\xrightarrow{q\to 1} (1-\frac{x^{(r)}_i}{x^{(r)}_j})^\b(1-\frac{x^{(r)}_j}{x^{(r)}_i})^\b,
\ee
\be
\frac{((qt)^{1/2}x^{(r+1)}_j/x^{(r)}_i;q)_{\infty}}
{((q/t)^{1/2}x^{(r+1)}_j/x^{(r)}_i;q)_{\infty}} \xrightarrow{q\to 1} (1-\frac{x^{(r+1)}_j}{x^{(r)}_i})^{-\b}.
\ee
From the above equations, we can reproduce the $A_n$ Selberg average. Besides, when $n=1$, the cross term in \eqref{q-integral} vanishes. By setting  notations agree, we can obtain the q-deformed $A_1$ case introduced in \cite{albionAFLTtypeSelbergIntegrals2021}.

We mainly consider the case where $\b$ is set to 1 in this paper. When $\b=1$, Macdonald polynomials $M_\l(x;q,t)(t=q^\b)$ reduce to the Schur functions $\chi_\l(x)$. The Schur functions are a crucial family of classical symmetric functions, which have the following definition \cite{macdonald1995}
\begin{equation}
\chi_{\l}(x)=
\frac{\det_{1\leq i,j\leq n}\big(x_i^{\l_j+n-j}\big)}{\Delta(x)},
\end{equation}
where $\Delta(x)$ is the Vandermonde products $\Delta(x)\equiv \prod_{1\leq i<j\leq n}(x_i-x_j).$

We define the $\b=1$ q-deformed $A_n$ Selberg average as follows:
\be\label{average}
\big\langle \mathscr{O}\big\rangle_{u_1,\dots,u_n,v}
^{N_1,\dots,N_n}\equiv
\frac{I^{A_n}_{N_1,\dots,N_n}
(\mathscr{O};u_1,\dots,u_n,v;1)}
{I^{A_n}_{N_1,\dots,N_n}(1;u_1,\dots,u_n,v;1)},
\ee
and 
\be\label{opm}
\big\langle \mathscr{O}\big\rangle_\pm\equiv\big\langle \mathscr{O}\big\rangle_{u_{1\pm},\dots,u_{n\pm},v_\pm}
^{N_{1\pm},\dots,N_{n\pm}}.
\ee

In \cite{albionAFLTtypeSelbergIntegrals2021} the Selberg integral formula for $n+1$ Schur polynomials is proposed. We now find a new q-deformed Selberg integral average formula, containing a product of $n+1$ Schur functions:
\begin{equation}\label{q-formula}
\begin{split}
&\bigg\langle \prod_{r=1}^{n+1} \chi_{\Yar{r}}\big[\xar{r}-\xar{r-1}\big]
\bigg\rangle_{u_1,\dots,u_n,v}
^{N_1,\dots,N_n}\\
=&\prod_{r=1}^{n+1}\prodp{\Yar{r}}\frac{1-q^{-N_r+N_{r-1}+i-j}}{1-q^{-\Yar{r}_i-Y^{\prime(r)}_j+i+j-1}}\prod_{1\leq r<s\leq n+1}\prod_{i=1}^{L_{\Yar{r}}}\prod_{j=1}^{L_{\Yar{s}}}\frac{1-q^{-\Yar{r}_i+\Yar{s}_j-A_{r,s}-j+i}}{1-q^{-A_{r,s}-j+i}}\\
\times&\prod_{1\leq r< s\leq n+1}\prodp{\Yar{r}}\frac{1-q^{-A_{r,s}+N_{s-1}-N_s+i-j}}{1-q^{-A_{r,s}-L_{\Yar{s}}+i-j}}\prodp{\Yar{s}}\frac{1-q^{-A_{r,s}+N_{r}-N_{r-1}-i+j}}{1-q^{-A_{r,s}+L_{\Yar{r}}-i+j}}.
    \end{split}
\end{equation}
where $N_0=x^{(0)}=0$ and $N_{n+1}=-v,\;x^{(n+1)}=-[v]_q$, and with the plethystic notation
\be
p_k[X-Y]=p_k[X]-p_k[Y],\quad p_k(X)=\sum_{i\geq1}x_i^k.
\ee
There are shorthand notations,
\be
A_r=u_r+\dots+u_n+N_r-N_{r-1}+n+1,\quad 1\leq r \leq n+1,
\ee
and for any $r$ and $s$,
\be\label{A}
    A_{r,s}=A_r-A_s=\sum_{i=r}^{s-1}u_i-\sum_{i=s}^{r-1}u_i+N_r-N_{r-1}-N_s+N_{s-1},
\ee
in particular, $A_{r,r}=0$. $l(\Yar{r})$ is the length of the partition $\Yar{r}$. $L_{\Yar{r}}$ 
\textup{(}$1\leqslant r\leqslant n+1$\textup{)}
is an arbitrary non-negative integer such that $L_{\Yar{r}}\geqslant l(\Yar{r})$. The proof of \eqref{q-formula} is given in appendix \ref{hint}. This new formula will be used in the direct proof of 5D AGT conjecture at $\b=1$ later.

\section{q-deformed conformal blocks}
In this section, we calculate the q-deformed conformal blocks corresponding to the 5D partition function. Conformal blocks can be used to build the correlation function in Toda field theory. The construction of conformal blocks in Toda theory involves the use of W-algebra associated with the underlying Lie algebra of the Toda system \cite{comanTodaConformalBlocks2020}. The Toda correlator can be represented by a multi-point function \cite{fateevCorrelationFunctionsConformal2007,fateevCorrelationFunctionsConformal2009}. 
In the case of $SU(N)$, the four point function corresponding to the conformal blocks can be written as Dotsenko-Fateev integral, which can be represented as a double Selberg average \cite{dotsenkoConformalAlgebraMultipoint1984,itoyamaMethodGeneratingQExpansion2010,Mironov:2010zs}.

Next, we deduce the q-deformed conformal blocks following the procedure for 4D case in \cite{itoyamaMethodGeneratingQExpansion2010,zhangSelbergIntegralSU2011}. The q-deformation of the double average can be expressed using the properties of q-deformation. The derivation is parallel to the 4D case, and the necessary changes are to replace the factors and integrals in 4D double average with their $q$-counterparts. 
As indicated by \eqref{q-shift} and \eqref{limit}, we can replace the following power-like factors with their product forms:
\begin{equation}
\label{rb}
(1-x)^{b} \rightarrow (x;q)_b,\quad b\in \mathbb{C}.
\end{equation}
According to this rule, the double average derived from Dotsenko-Fateev integral in \cite{zhangSelbergIntegralSU2011} becomes
\be\label{double}
\left\langle \!\!\! \left\langle
\prod_{r=1}^{n} \left\{ \prod_{i=1}^{N_r} (\cq x_{i}^{(r)};q)_{v_{r-}} 
\prod_{j=1}^{\tilde{N_r}} (\cq y_{i}^{(r)};q)_{v_{r+}}
\right\}
\prod_{r,s=1}^{n}\prod_{i=1}^{N_r}\prod_{j=1}^{\tilde{N_s}}\left(\cq x_{i}^{(r)} y_{j}^{(s)};q \right)_{\b} ^{C_{rs}}
\right\rangle_{+} \right\rangle_{-} \;,
\ee
where $<>_{\pm}$ are defined by \eqref{opm}, and $C_{rs}$ is the Cartan matrix
\[ C_{rs}=\begin{cases}
2 & r=s\\
-1 & r=s \pm 1 \\
0 & |r-s| > 1 \; .
\end{cases} \]
We divide the integral of \eqref{double} into three parts and change them separately into exponential form. The first part is
\be
\begin{split}
    \prod_{r=1}^{n} \prod_{i=1}^{N_r}(\cq x_{i}^{(r)};q)_{v_{r-}} &= \exp\left( \sum_{r=1}^{n} \sum_{i=1}^{N_r}\ln\left(\frac{\prod_{k\geq0}1-\cq\xar{r}_iq^k}{\prod_{k\geq0}1-\cq\xar{r}_{i}q^{k+v_{r-}}}\right)\right)\\
&=\exp\left(-\sum_{r=1}^{n} \sum_{i=1}^{N_r}\sum_{m=1}^{\infty}\frac{\cq^m {(x_{i}^{(r)})}^m}{m}\left(\sum_{k\geq0}q^{km}-\sum_{k\geq0}q^{(k+v_{r-})m}\right)\right)\\
&=\exp\left(-\sum_{m=1}^{\infty}[\beta]_{q^m} \frac{\cq^m }{m} \sum_{r=1}^{n} p_{m}^{(r)} \frac{[v_{r_-}]_{q^m}}{[\beta]_{q^m}} \right)\\
&=\exp\left(-\sum_{m=1}^{\infty}[\beta]_{q^m} \frac{\cq^m }{m} \sum_{r=1}^{n+1} z_{m}^{(r)} \frac{[v_{r_-}]^{\prime}_{q^m}}{[\beta]_{q^m}} \right).
\end{split}
\ee
In the second equation, we perform Taylor expansion. In the third equation, we define
\be
p_m^{(r)}:=\sum_i(x^{(r)}_i)^m,\quad [v_{r_{\pm}}]_{q^m}=\dfrac{\ \ 1-q^{m v_{r_{\pm}}}}{1-q^m}=\sum_{k\geq0}
q^{km}-\sum_{k\geq0}q^{(k+v_{r\pm})m}.
\ee
In the final step, we use the following notations
\be
\begin{split}
&z_m^{(r)}\equiv p_m^{(r)}-p_m^{(r-1)}, \quad  [v_{r-}]_{q^m}^{\prime} \equiv-\sum_{s=1}^{r-1} [v_{s-}]_{q^m} , \\
&x^{(0)}=x^{(n+1)}=N_{0-}=N_{(n+1)-}=0 .
\end{split}
\ee
In the same way, we obtain the second part
\be
\begin{split}
\prod_{r=1}^{n}\prod_{j=1}^{\tilde{N_r}} 
(\cq y_{i}^{(r)};q)_{v_{r+}} 
=\exp\left( - \sum_{m=1}^{\infty}[\beta]_{q^m} \frac{\cq^m }{m} \sum_{r=1}^{n}\tilde p_{m}^{(r)}
\frac{[v_{r_+}]_{q^m}}{[\beta]_{q^m}} \right)\\
=\exp\left( -\sum_{m=1}^{\infty}[\beta]_{q^m} \frac{\cq^m }{m} \sum_{r=1}^{n+1} \tilde z_{m}^{(r)}
\frac{[v_{r_+}]^{\prime}_{q^m}}{[\beta]_{q^m}} \right).    
\end{split}
\ee
Here we use the notations
\be
\begin{split}
&\tilde p_m^{(r)}:=\sum_i(y^{(r)}_i)^m,\quad \tilde z_m^{(r)}:=\tilde p_m^{(r)}-\tilde p_m^{(r-1)},\\
&[v_{(n+1-r)+}]_{q^m}^{\prime} \equiv \sum_{s=1}^{r} [v_{(n+1-s)+}]_{q^m},\\
&y^{(0)}=y^{(n+1)}=N_{0+}=N_{(n+1)+}=0 .
\end{split}
\ee
Similarly, the third part becomes
\be
\begin{split}
   & \prod_{r,s=1}^{n}\prod_{i=1}^{N_r}\prod_{j=1}^{\tilde{N_s}}
\left(\cq x_{i}^{(r)} y_{j}^{(s)} ;q\right)_{\b}^{C_{rs}}\\
=& \exp\left( -\sum_{r,s=1}^{n}C_{rs}\sum_{i=1}^{N_r}\sum_{j=1}^{\tilde{N_s}}\sum_{m=1}^{\infty}\frac{\cq^m {(x_{i}^{(r)})}^m {(y_{j}^{(s)})}^m}{m} \left(\sum_{k\geq0}q^{km}-\sum_{k\geq 0}q^{(k+\b)m}\right)\right)\\
=&\exp\left(\sum_{m=1}^{\infty}[\beta]_{q^m} \frac{\cq^m }{m}
\Bigg[ 2 \sum_{r=1}^{n} p_m^{(r)} \tilde{p}_m^{(r)}
-\sum_{r=2}^{n} p_m^{(r)} \tilde{p}_m^{(r-1)}
-\sum_{r=1}^{n-1} p_m^{(r)} \tilde{p}_m^{(r+1)}\Bigg ]\right)\\
=&\exp\left( -  \sum_{r=1}^{n+1} \sum_{m=1}^{\infty}[\beta]_{q^m} \frac{ \cq^m }{m}  z_m^{(r)} \tilde{z}_m^{(r)}
\right).
\end{split}
\ee
Combining the results of the above three parts and applying Cauchy formula for the Macdonald polynomials \cite{mironovProvingAGTConjecture2012}
\be
\exp\Big( \sum\limits_{m=1}^{\infty} \dfrac{[\beta]_{q^m}}{m} p_{m} {\widetilde p}_{m} \Big)=\sum\limits_{Y} \frac{C_{Y}}{C_{Y}^{\prime}}M_{Y}(p_{m}) M_{Y}( {\widetilde p}_{m}),
\ee
where
\be 
C_{Y}^{\prime}=\prod\limits_{(i,j)\in Y}\,[\beta \textrm{Arm}_{Y}(i,j)+
\textrm{Leg}_{Y}(i,j) +\beta ]_q,\ \ \ C_{Y}=\prod\limits_{(i,j)\in Y}\,[\beta \textrm{Arm}_{Y}(i,j)+\textrm{Leg}_{Y}(i,j) +1 ]_q,
\ee
the integral \eqref{double} becomes
\begin{equation}\label{5D product}
\begin{split}
       & \left\langle \!\!\!\left\langle\text{exp} \Bigg\{ -   \sum_{m=1}^{\infty} [\beta]_{q^m}\frac{\cq^m}{m}
\sum_{r=1}^{n+1}\Big[ (z_m^{(r)} + \frac{[v_{r_+}]^{\prime}_{q^m}}{[\beta]_{q^m}})(\tilde{z}_m^{(r)}+ \frac{[v_{r_-}]^{\prime}_{q^m}}{[\beta]_{q^m}})
-\frac{[v_{r_+}]^{\prime}_{q^m}}{[\beta]_{q^m}} \frac{[v_{r_-}]^{\prime}_{q^m}}{[\beta]_{q^m}}
\Big]
\Bigg\}\right\rangle_{+} \right\rangle_{-} 
\\
=&\left\langle \!\!\! \left\langle \prod_{r=1}^{N} (1-\cq)^{[v_{r_+}]^{\prime}_{q^m} [v_{r_-}]^{\prime}_{q^m} / [\beta]_{q^m}}
\sum_{\vec{Y}}\prod_{r=1}^{n+1}
 \cq^{\lvert \vec{Y} \rvert} \frac{C_{Y^{(r)}}}{C_{Y^{(r)}}^{\prime}}
M_{Y^{(r)}} (-z_m^{(r)} - \frac{[v_{r_+}]^{\prime}_{q^m}}{[\beta]_{q^m}}) M_{Y^{(r)}} (\tilde{z}_m^{(r)}+ \frac{[v_{r_-}]^{\prime}_{q^m}}{[\beta]_{q^m}}) \;\Bigg\}\right\rangle_{+} \right\rangle_{-} 
\\
=&\prod_{r=1}^{N} (1-\cq)^{[v_{r_+}]^{\prime}_{q^m} [v_{r_-}]^{\prime}_{q^m} / [\beta]_{q^m}}\\
&\sum_{\vec{Y}} \cq^{\lvert \vec{Y} \rvert} \frac{C_{Y^{(r)}}}{C_{Y^{(r)}}^{\prime}}\left\langle\prod_{r=1}^{n+1}M^{([\beta]_{q^m})}_{Y^{(r)}} (-z_m^{(r)} - \frac{[v_{r_+}]^{\prime}_{q^m}}{[\beta]_{q^m}})\right\rangle_{+}\left\langle\prod_{r=1}^{n+1}
M^{([\beta]_{q^m})}_{Y^{(r)}} (\tilde{z}_m^{(r)}+ \frac{[v_{r_-}]^{\prime}_{q^m}}{[\beta]_{q^m}})
\right\rangle_{-} \;.
\end{split}
\end{equation}
When $\b=1$, Macdonald polynomials reduce to Schur polynomials. For convenience of the subsequent proof, we use $v_{r+}=v_+\delta_{r1}$, $v_{r-}=v_-\delta_{rn}$ and redefine $N_{0+}=-v_+,\;\;x^{(0)}=-[v_+]_q$ and $N_{(n+1)-}=-v_-,\;\;y^{(n+1)}=-[v_-]_q$. Absorbing the prefactor we can rewrite the result of \eqref{5D product} as
\be\label{q-cb}
\sum_{\vec{Y}} \cq^{\lvert \vec{Y} \rvert} \left<\prod_{r=1}^{n+1}\chi_{Y^{(r)}}\left[x^{(r-1)}-x^{(r)}\right]\right>_+ 
\left<\prod_{r=1}^{n+1}\chi_{Y^{(r)}}\left[y^{(r)}-y^{(r-1)}\right]\right>_-.
\ee
We now complete the calculation of the q-deformed conformal blocks.

\section{Proof for 5D AGT at $\b=1$}
In this section, we prove 5D $A_n$ AGT conjecture in the special case of $\b=1$. The method of proof is similar to the 4D case introduced in \cite{yuanProofA_AGT2023}. The main difference is that this time we make use of the q-deformed $A_n$ Selberg integral formula \eqref{q-formula} proposed in section \ref{Selberg}.

5D AGT conjecture implies that the 5D partition function \eqref{5d partition}
and the q-deformed correlator \eqref{q-cb} are the same (with some prefactors), that is to say
\be
\label{5d eq}
\begin{split}
\left<\prod_{r=1}^{n+1}\chi_{Y^{(r)}}\left[x^{(r-1)}-x^{(r)}\right]\right>_+ 
\left<\prod_{r=1}^{n+1}\chi_{Y^{(r)}}\left[y^{(r)}-y^{(r-1)}\right]\right>_-
\\
=\prod_{r,s=1}^{n+1}\frac{\cF_{Y^{(s)}} (\mu_r+a_s,q)\cF_{Y^{(s)}} (\mu_{n+1+r}+a_s,q^{-1})}{\cG_{Y^{(r)},Y^{(s)}}(a_r-a_s,q)\cG_{Y^{(s)},Y^{(r)}}(a_s-a_r+1-\b,q^{-1})}.
\end{split}
\ee

The proof follows in two steps: first, we apply the Selberg integral formula to the two integrals on the left side of \eqref{5d eq} separately. Second, we use the following lemma to transform the results of integrals to the right hand side of \eqref{5d eq}.

\textbf{Lemma 1}
\begin{equation}\label{lemma}
\begin{split}
    \prod_{i=1}^{L_Y}\prod_{j=1}^{L_W}\frac{1-q^{x+Y_i-W_j+j-i}}{1-q^{x+j-i}}
        &\prodp{Y}\frac{1}{1-q^{x+L_W-i+j}}
        \prodp{W}\frac{1}{1-q^{x-L_Y+i-j}}\\
        =&\left.\frac{1}{\cG_{Y,W}(-x,q^{-1})\cG_{W,Y}(x,q)}\right|_{\beta=1}.
    \end{split}
\end{equation}
This lemma is proved in Appendix \ref{proof_lemma}.
The left hand side of it corresponds to the terms in the integral formula.

Before we use the q-deformed $A_n$ Selberg integral formula \eqref{q-formula}, we notice that the form of the first integral on the left side of the \eqref{5d eq} is slightly different from the left side of the formula.
We can change the form of the product of Schur polynomials 
\begin{equation}\label{first}
    \begin{split}
        &\prod_{r=1}^{n+1}\chi_{W^{(r)}}\left[z^{(r)}-z^{(r-1)}\right]=\prod_{r=1}^{n+1}\chi_{Y^{(n+2-r)}}\left[x^{(n+1-r)}-x^{(n+1-(r-1))}\right]\\
        =&\prod_{r=1}^{n+1}\chi_{Y^{(n+2-r)}}\left[x^{((n+2-r)-1)}-x^{(n+2-r)}\right]=\prod_{r=1}^{n+1}\chi_{Y^{(r)}}\left[x^{(r-1)}-x^{(r)}\right].
    \end{split}
\end{equation}
In the first equation, we use the relationships $z^{(r)}=x^{(n+1-r)}$ and $W^{(r)}=Y^{(n+2-r)}$. In the last step, the products of items from $n+2-r=n+1$ to $n+2-r=1$ are converted to products of items form $r=1$ to $r=n+1$. When we integrate the original and the inverse form of products, all items in two integrals could be equated respectively by similar procedure. Thus,
\begin{equation}
  \left<\prod_{r=1}^{n+1}\chi_{Y^{(r)}}\left[x^{(r-1)}-x^{(r)}\right]\right>_{u_{1+},...,u_{n+},v+}^{N_{1+},...,N_{n+}}=\left<\prod_{r=1}^{n+1}\chi_{W^{(r)}}\left[z^{(r)}-z^{(r-1)}\right]\right>_{u_{n+},...,u_{1+},v+}^{N_{n+},...,N_{1+}}.
\end{equation}

Applying the lemma to the result of the integral, we obtain the first part
\begin{equation}\label{x-part}
\begin{split}
& \left<\prod_{r=1}^{n+1}\chi_{Y^{(r)}}\left[x^{(r-1)}-x^{(r)}\right]\right>_+
=\left<\prod_{r=1}^{n+1}\chi_{W^{(r)}}\left[z^{(r)}-z^{(r-1)}\right]\right>_{u_{n+},...,u_{1+},v+}^{N_{n+},...,N_{1+}}\\
=&\prod_{r=1}^{n+1}\frac{\cF_{W^{(r)}}(N_{(n-r+2)+}-N_{(n+1-r)+},q)}{G_{W^{(r)},W^{(r)}}(0,q^{-1})} \\
&\prod_{1\leq r<s\leq n+1}\bigg\{\frac{\cF_{W^{(s)}}(A^+_{n-s+2,n-r+2}+N_{(n-r+2)+}-N_{(n-r+1)+},q^{-1})}{\cG_{W^{(r)},W^{(s)}}(A^+_{n-r+2,n-s+2},q)}\\
&\qquad\qquad\times\frac{\cF_{W^{(r)}}(A^+_{n-r+2,n-s+2}+N_{(n-s+2)+}-N_{(n-s+1)+},q)}{\cG_{W^{(s)},W^{(r)}}(A^+_{n-s+2,n-r+2},q^{-1})}\bigg\}\\
=&\prod_{r=1}^{n+1}\frac{\cF_{\Yar{r}}(N_{r+}-N_{(r-1)+},q)}{\cG_{\Yar{r},\Yar{r}}(0,q^{-1})} \\
&\prod_{1\leq r<s\leq n+1}\frac{\cF_{Y^{(r)}}(A^+_{r,s}-N_{(s-1)+}+N_{s+},q^{-1})\cF_{Y^{(s)}}(A^+_{s,r}-N_{(r-1)+}+N_{r+},q)}{\cG_{Y^{(r)},Y^{(s)}}(A^+_{r,s},q^{-1})\cG_{Y^{(s)},Y^{(r)}}(A^+_{s,r},q)}.
\end{split}
\end{equation}
In the first step, we use the definition of $\cF_Y$. In the second step, we use $Y$ instead of $W$.

Similarly, we can obtain the second part
\begin{equation}\label{y-part}
\begin{split}
&\left<\prod_{r=1}^{n+1}\chi_{Y^{(r)}}\left[y^{(r)}-y^{(r-1)}\right]\right>_-
=\prod_{r=1}^{n+1}\frac{\cF_{\Yar{r}}(N_{(r-1)-}-N_{r-},q)}{\cG_{\Yar{r},\Yar{r}}(0,q^{-1})} \\
&\times\prod_{1\leq r<s\leq n+1}\frac{\cF_{Y^{(s)}}(A^-_{r,s}+N_{(r-1)-}-N_{r-},q^{-1})\cF_{Y^{(r)}}(A^-_{s,r}+N_{(s-1)-}-N_{s-},q)}{\cG_{Y^{(r)},Y^{(s)}}(A^-_{s,r},q)\cG_{Y^{(s)},Y^{(r)}}(A^-_{r,s},q^{-1})},
\end{split}
\end{equation}
where $A^+$ and $A^-$ respectively correspond to parameters in integrals for $x$ and $y$. According to the form of $\cF_Y(x,q)$, we find that 
%\be
%\begin{split}
% &\cF_{Y^{(s)}} (-(\mu_r+a_s),q)\cF_{Y^{(s)}} (-(\mu_{n+1+r}+a_s),q^{-1})\\
%=&\sqrt{\cF_{Y^{(s)}} (-(\mu_r+a_s),q)\cF_{Y^{(s)}} (-(\mu_r+a_s),q^{-1})\cF_{Y^{(s)}} (-(\mu_{n+1+r}+a_s),q)\cF_{Y^{(s)}} (-(\mu_{n+1+r}+a_s),q^{-1})}   
%\end{split}
%\ee
\be\label{F relation}
\cF_{Y^{(s)}} (x,q)=\prod_{(i,j)\in Y^{(s)}}-q^{x-j+1}t^{i-1} \cF_{Y^{(s)}}(x,q^{-1}) .
\ee
Considering the definition of $A_{r,s}$ given in \eqref{A}, we give the relations of parameters
\begin{equation}
    \begin{split}
        \mu_{s}+a_r&=\sum_{i=r}^{s-1}u_{i+}-\sum_{i=s}^{r-1}u_{i+}+N_{r+}-N_{(r-1)+},\\
        a_r-a_s&=\sum_{i=r}^{s-1}u_{i+}-\sum_{i=s}^{r-1}u_{i+}+N_{r+}-N_{(r-1)+}-N_{s+}+N_{(s-1)+},
    \end{split}
\end{equation}
\begin{equation}
    \begin{split}
        \mu_{s+n+1}+a_r&=-\left(\sum_{i=r}^{s-1}u_{i-}-\sum_{i=s}^{r-1}u_{i-}+N_{r-}-N_{(r-1)-}\right) ,\\
        a_r-a_s&=-\left(\sum_{i=r}^{s-1}u_{i-}-\sum_{i=s}^{r-1}u_{i-}+N_{r-}-N_{(r-1)-}-N_{s-}+N_{(s-1)-}\right) ,
    \end{split}
\end{equation}
where $r$ and $s$ run from $1$ to $n+1$. Setting notations to be consistent, multiplying the terms on both sides of \eqref{x-part} and \eqref{y-part}, and using \eqref{F relation}, we can get \eqref{5d eq}. Thus, we finish the proof of 5D $A_n$ AGT conjecture at $\b=1$.

\section{Conclusion}
In this paper, we propose and prove a q-deformed $A_n$ Selberg average formula with $n+1$ Schur polynomials. Based on this new formula, we give a proof of 5D $A_n$ AGT conjecture at $\b=1$. 

Recently, multiple extension forms of Selberg integral in the context of AGT correspondence have regained some progress, including the AFLT Selberg integral over a product of Jack polynomials and elliptic generalization of the AFLT integral containing elliptic interpolation functions \cite{albaCombinatorialExpansionConformal2011,albionAFLTtypeSelbergIntegrals2021,albionEllipticMathrm_n2023}, which correspond to 4D and 6D gauge theories respectively. The new q-deformed $A_n$ Selberg average formula given in this article corresponds to the 5D gauge theory.

In addition, this work generalizes the direct proof for 4D $A_n$ AGT conjecture at $\b=1$ introduced in \cite{yuanProofA_AGT2023}. We confirm that the 4D $A_n$ AGT correspondence can be extended to an equivalence between the q-deformed conformal blocks and the 5D instanton partition functions. 
In the lifted gauge theory, the derivation of the 5D partition function is accomplished by implementing appropriate q-transformation to all factors of the 4D partition function. We generalize the $A_n$ Selberg integral to its q-deformed counterpart, which can be used to derive explicit representation of q-deformed conformal blocks. We also verify the rationality of the new integral form. Taking the $q \rightarrow 1$ limit, the new integral reduces to the original one. Following the procedure introduced in \cite{mironovProvingAGTConjecture2012,zhangSelbergIntegralSU2011}, we complete the calculation of the q-deformed conformal blocks. We utilize a q-deformed $A_n$ Selberg average formula and a lemma to prove 5D $A_n$ AGT conjecture in the special case of $\b=1$. In the future, we will work on the general $\b$ case.

\acknowledgments
The authors thank S. Ole Warnaar, Yutaka Matsuo, Jean-Emile Bourgine, and Katsushi Ito for helpful comments.  KZ (Hong Zhang) is supported by a classified fund from Shanghai city.

\appendix
\section{Macdonald polynomials}

Macdonald polynomials, denoted as $ M_{\lambda}(x; q, t)$, are a family of symmetric polynomials indexed by partitions $\lambda$. They are defined over a set of variables $x = (x_1, x_2, \ldots, x_n)$ and depend on two parameters $q$ and $t$ \cite{macdonald1995}.

Macdonald polynomials are eigenfunctions of a certain difference operator, which is a $q$-analogue of the usual differential operator. These polynomials generalize several well-known families of polynomials in symmetric function theory.
The Macdonald difference operator is defined as
\begin{align}
\begin{split}
    D_{n}^{(r)}(x;q,t)=\sum_{I}\prod_{\substack{i \in I\\ j\not\in I}}t^{r(r-1)/2}\frac{tx_{i}-x_{j}}{x_{i}-x_{j}}\prod_{i\in I}T_{q,x_{i}},\quad r=0,1,\ldots,n,
\end{split}
\end{align}
where $T_{q,x_{i}}$ is a q-shift operator, defined as $T_{q,x_{i}}x_j=q^{\delta_{ij}} x_j$. We consider the eigenfunctions of these operators
\begin{align}
    D_{n}^{(r)}(x;q,t)M_{\lambda}(x;q,t)=e_{r}(q,t)M_{\lambda}(x;q,t),\quad r=0,1,\ldots,n.
\end{align}
The eigenfunctions $M_{\lambda}(x;q,t)$ are imposed to satisfy the renormalization condition,
\begin{equation}
    M_{\lambda}(x; q,t)=m_{\lambda}(x)+\sum_{\mu<\lambda}u_{\lambda\mu}(q,t)m_{\mu}(x),
\end{equation}
which are called Macdonald polynomials. 

Using the power sum $p_k(x)=\sum_i x_i^k$, the explicit forms of the first several Macdonald polynomials are listed as follows
\begin{align*}
&M_{1}=p_{1}, \\
&M_{2}=\frac{(1-t)(1+q)}{(1-t q)} \frac{p_{1}^{2}}{2}+\frac{(1+t)(1-q)}{(1-t q)} \frac{p_{2}}{2}, \\
&M_{11}=\frac{p_{1}^{2}}{2}-\frac{p_{2}}{2}, \\
&M_{3}=\frac{(1+q)\left(1-q^{3}\right)(1-t)^{2}}{(1-q)(1-t q)\left(1-t q^{2}\right)} \frac{p_{1}^{3}}{6}+\frac{\left(1-t^{2}\right)\left(1-q^{3}\right)}{(1-t q)\left(1-t q^{2}\right)} \frac{p_{1} p_{2}}{2}+\frac{(1-q)\left(1-q^{2}\right)\left(1-t^{3}\right)}{(1-t)(1-t q)\left(1-t q^{2}\right)} \frac{p_{3}}{3}, \\
&M_{21}=\frac{(1-t)(2 q t+q+t+2)}{1-q t^{2}} \frac{p_{1}^{3}}{6}+\frac{(1+t)(t-q)}{1-q t^{2}} \frac{p_{1}}{2}-\frac{(1-q)\left(1-t^{3}\right)}{(1-t)\left(1-q t^{2}\right)} \frac{p_{3}}{3},\\ &M_{111}=\frac{p_{1}^{3}}{6}-\frac{p_{1} p_{2}}{2}+\frac{p_{3}}{3}.
\end{align*}

One of the key features of Macdonald polynomials is taking limits on them can yield other symmetric polynomials. After taking the $t = q$ limit, the polynomials become functions of only $x$ and are independent of q, called Schur functions $\chi_\l(x)$. When $q$ tends to 1 while setting $t=q^{\b}$, Macdonald polynomials become Jack functions $P_\l(x)$. 
%Schur functions and Jack functions are both mentioned in the main text.

\section{Proof of the q-deformed Selberg integral formula}\label{hint}
Here we provide a proof for the q-deformed Selberg average formula \eqref{q-formula} over $n+1$ Schur polynomials proposed in section \ref{Selberg}. The formula reads
\begin{equation}\label{formula}
\begin{split}
&\bigg\langle \prod_{r=1}^{n+1} \chi_{\Yar{r}}\big[\xar{r}-\xar{r-1}\big]
\bigg\rangle_{u_1,\dots,u_n,v}
^{N_1,\dots,N_n}\\
=&\prod_{r=1}^{n+1}\prodp{\Yar{r}}\frac{1-q^{-N_r+N_{r-1}+i-j}}{1-q^{-\Yar{r}_i-Y^{\prime(r)}_j+i+j-1}}\prod_{1\leq r<s\leq n+1}\prod_{i=1}^{L_{\Yar{r}}}\prod_{j=1}^{L_{\Yar{s}}}\frac{1-q^{-\Yar{r}_i+\Yar{s}_j-A_{r,s}-j+i}}{1-q^{-A_{r,s}-j+i}}\\
\times&\prod_{1\leq r< s\leq n+1}\prodp{\Yar{r}}\frac{1-q^{-A_{r,s}+N_{s-1}-N_s+i-j}}{1-q^{-A_{r,s}-L_{\Yar{s}}+i-j}}\prodp{\Yar{s}}\frac{1-q^{-A_{r,s}+N_{r}-N_{r-1}-i+j}}{1-q^{-A_{r,s}+L_{\Yar{r}}-i+j}}.
    \end{split}
\end{equation}
The q-deformed Selberg average is defined in \eqref{average}. Firstly we calculate
\be
\label{integ}
I^{A_n}_{N_1,...,N_n}\left(\prod_{r=1}^{n+1}\chi_{\Yar{r}}[\xar{r}-\xar{r-1}],u_1,...,u_n,v;1\right).
\ee
Following the method in \cite{albionAFLTtypeSelbergIntegrals2021}, we denote the complex Schur function
\be
S^{(n)}(x;z):=\frac{\det_{1\leq i,j\leq n}\big(x_i^{z_j}\big)}{\Delta(x)},
\ee
which has the following properties 
\be\label{complexschur1qq}
S^{(n)}(1,q,\dots,q^{n-1};z)=S^{(n)}\left([n]_q;z\right)=\prod_{i=1}^{n}q^{(i-1)(z_i-n+i)}\prod_{1\leq i<j\leq n}\frac{1-q^{z_i-z_j}}{1-q^{j-i}}.
\ee
\begin{equation}
\chi_{Y}(x_1,\dots,x_n)=
S^{(n)}(x_1,\dots,x_n;Y_1+n-1,Y_2+n-2,\dots,Y_n),
\end{equation}
where $Y$ is a partition, $Y=(Y_1,\dots,Y_n)$. And Schur polynomials satisfy \cite{macdonald1995},
\be
\begin{split}
\label{schur1qq}
\chi_Y(1,q,\dots,q^{n-1})=\chi_Y\left([n]_q\right)=&\prod_{i\geq 1}q^{(i-1)Y_i}\prodp{Y}\frac{1-q^{n-i+j}}{1-q^{Y_i+Y'_j-i-j+1}}\\
=&\prod_{i\geq 1}q^{(i-1)Y_i}\prod_{1\leq i<j\leq n}\frac{(q^{j-i+1};q)_{Y_i-Y_j}}{(q^{j-i};q)_{Y_i-Y_j}}.
\end{split}
\ee
We can rewrite \eqref{integ} as,
{\small
\be
\begin{split}
&\frac{q^{f(a_n)}}{N_1!\cdots N_n!(2\pi\iup)^{N_1+\cdots+N_n}} \Int_{\mathbb{T}^{N_1+\cdots+N_n}} \prod_{r=1}^{n+1} \chi_{\Yar{r}}\big[\xar{r}-\xar{r-1}\big]
\prod_{i=1}^{N_n}\frac{\prod_{k=-\infty}^{+\infty}(\xar{n}_i-q^{a_n+k})}
{\prod_{k=-\infty}^{0}(\xar{n}_i-q^k)\prod_{k=1}^{+\infty}(\xar{n}_i-q^{v+k})}\\
\times&\prod_{r=1}^{n}\prod_{i=1}^{N_r}(\xar{r}_i)^{u_r}\prod_{r=1}^n \prod_{1\leqslant i<j\leqslant N_r} (\xar{r}_i-\xar{r}_j)^2\prod_{r=1}^{n-1}\prod_{i=1}^{N_r}\prod_{j=1}^{N_{r+1}}(\xar{r}_i-\xar{r+1}_j)^{-1}
\dup x^{(1)}\cdots \dup x^{(n)},    
\end{split}
\ee
}
where $f(a_n)$ is a function of parameter $a_n$. Since this factor will be eliminated at the final average, we do not have to write down its explicit form. By using theorem 5.3 and the q-deformation of theorem 5.5 in \cite{albionAFLTtypeSelbergIntegrals2021} (in our case there might be an ambiguity in choosing the contour), it can be simplified to
\be
\label{result-1}
\begin{split}
\prod_{r=1}^n(-1)^{\binom{N_r}{2}}&\prod_{i=1}^{N_n}q^{-i(v+N_n)}q^{f(a_n)-N_n|\Yar{n+1}|}S^{(N_{n})}\left(-[-N_n]_q;z^{(n)}+u_n\right)\chi_{\Yar{n+1}}(-[v+N_n]_q)\\
&\times\prod_{i=1}^{N_n}\left\{\frac{(q^{-z^{(n)}_i-u_n+N_{n+1}-N_n-1};q^{-1})_{N_{n+1}-N_n}}{(q^{-N_n+i-1};q^{-1})_{N_{n+1}-N_n}}\prod_{j\geq 1}\frac{1-q^{-z^{(n)}_i-u_n+\Yar{n+1}_j+N_{n+1}-N_n-j}}{1-q^{-z^{(n)}_i-u_n+N_{n+1}-N_n-j}}\right\},
\end{split}
\ee
where $z^{(n)}$ is defined recursively by,
\ba
&(z^{(r)}_1,\dots,z^{(r)}_{N_r})\\
=&(z^{(r-1)}_1+u_{r-1},\dots,z^{(r-1)}_{N_{r-1}}+u_{r-1},\Yar{r}_1+N_r-N_{r-1}-1,\dots,\Yar{r}_{N_r-N_{r-1}}),
\ea
for $1\leq r\leq n+1$ and $z^{(0)}$ empty. 

The main difference between our case and that in \cite{albionAFLTtypeSelbergIntegrals2021} is the value of $\xar{n+1}$. In four dimension, $\xar{n+1}=1+1+1\cdots+1$, while here,
\be
\xar{n+1}=-[v]_q=-[-N_{n+1}]_q=-\frac{1-q^{-N_{n+1}}}{1-q}=\frac{q^{-1}(1-q^{-N_{n+1}})}{1-q^{-1}}=q^{-1}+q^{-2}+\cdots+q^{-N_{n+1}},
\ee
which indeed denotes the pole in
\be
\prod_{i=1}^{N_n}\frac{\prod_{k=-\infty}^{+\infty}(\xar{n}_i-q^{a_n+k})}
{\prod_{k=-\infty}^{0}(\xar{n}_i-q^k)\prod_{k=1}^{+\infty}(\xar{n}_i-q^{-N_n+k})}.
\ee
In the case of $v=-N_{n+1}$ a negative integer and $u_n$ an integer, the above results can be strictly proved by applying following relation derived from \eqref{complexschur1qq} and \eqref{schur1qq}
\ba
\label{rec-q-schur}
\prod_{i=1}^{k}q^{i(\ell-k)}q^{-k|Y|}&S^{(k)}(-[-k]_q;z)\chi_{Y}[-[-\ell+k]_q]\prod_{i=1}^{k}\prod_{j=1}^{\ell-k}\frac{1-q^{-z_i+Y_j+\ell-k-j}}{1-q^{-k-j+i}}\\
=&\begin{cases}
    S^{(\ell)}\left(-[-\ell]_q;(z,Y_1+\ell-k-1,...,Y_{\ell-k})\right)\;&\text{if}\;l(Y)\leq \ell-k,\\
    0&\text{otherwise.}
\end{cases}
\ea
The non-integer $v$ and $u_n$ case can also be proved by analytical continuation. After recursion through \eqref{rec-q-schur}, \eqref{result-1} can be written as,
\ba
q^{f(a_n)}&\prod_{r=1}^{n}\left(\frac{(-1)^{\binom{N_r}{2}}q^{-N_r|\Yar{r+1}|}\prod_{i=1}^{N_r}q^{i(N_{r+1}-N_r)}}{\prod_{i=1}^{N_r}(q^{-N_r+i-1};q^{-1})_{N_{r+1}-N_r}}\right)\prod_{r=1}^{n+1}\chi_{\Yar{r}}[-[-N_{r}+N_{r-1}]_q]\\
&\times\prod_{1\leq r<s\leq n}\prod_{i=1}^{N_r-N_{r-1}}\prod_{j=1}^{N_s-N_{s-1}}(1-q^{-\Yar{r}_i+\Yar{s}_j-A_{r,s}-j+i})\\
&\times\prod_{r=1}^n\prod_{i=1}^{N_r-N_{r-1}}\left(\frac{(q^{-\Yar{r}_i-A_{r,n+1}+i-1};q^{-1})_{N_{n+1}-N_n}}{(q^{-N_n+i-1};q^{-1})_{N_{n+1}-N_n}}\prod_{j\geq 1}\frac{1-q^{-\Yar{r}_i+\Yar{n+1}_j-A_{r,n+1}-j+i}}{1-q^{-\Yar{r}_i-A_{r,n+1}-j+i}}\right).
\ea
Then we have,
\ba
&I^{A_n}_{N_1,...,N_n}(1,u_1,...,u_n,v;1)
=q^{-v(v-1)/2}\prod_{r=1}^{n}\left(\frac{(-1)^{\binom{N_r}{2}}\prod_{i=1}^{N_r}q^{i(N_{r+1}-N_r)}}{\prod_{i=1}^{N_r}(q^{-N_r+i-1};q^{-1})_{N_{r+1}-N_r}}\right)\\
&\times\prod_{1\leq r<s\leq n}\prod_{i=1}^{N_r-N_{r-1}}\prod_{j=1}^{N_s-N_{s-1}}(1-q^{-A_{r,s}-j+i})
\prod_{r=1}^n\prod_{i=1}^{N_r-N_{r-1}}\frac{(q^{-A_{r,n+1}+i-1};q^{-1})_{N_{n+1}-N_n}}{(q^{-N_n+i-1};q^{-1})_{N_{n+1}-N_n}}.
\ea
Now we can get the q-deformed Selberg average for $n+1$ Schur functions,
\ba
&\bigg\langle \prod_{r=1}^{n+1} \chi_{\Yar{r}}\big[\xar{r}-\xar{r-1}\big]
\bigg\rangle_{u_1,\dots,u_n,v}
^{N_1,\dots,N_n}=\prod_{r=1}^n\left(q^{-N_r|\Yar{r+1}|}\prod_{i=1}^{l_{\Yar{1}}}q^{-iu_r}\right)\\
\times&\prod_{r=1}^{n+1}\chi_{\Yar{r}}[-[-N_{r}+N_{r-1}]_q]\prod_{1\leq r<s\leq n+1}\prod_{i=1}^{L_{\Yar{r}}}\prod_{j=1}^{L_{\Yar{s}}}\frac{(1-q^{-\Yar{r}_i+\Yar{s}_j-A_{r,s}-j+i})}{(1-q^{-A_{r,s}-j+i})}\\
\times&\prod_{1\leq r< s\leq n+1}\prodp{\Yar{r}}\frac{1-q^{-A_{r,s}+N_{s-1}-N_s+i-j}}{1-q^{-A_{r,s}-L_{\Yar{s}}+i-j}}\prodp{\Yar{s}}\frac{1-q^{-A_{r,s}+N_{r}-N_{r-1}-i+j}}{1-q^{-A_{r,s}+L_{\Yar{r}}-i+j}}.
\ea
After isolating some trivial q-factor and using \eqref{schur1qq}, this result is equivalent to \eqref{formula}.

\section{Proof of Lemma 1}\label{proof_lemma}
Here we prove the lemma \eqref{lemma} in the main text.

~\\\textbf{Lemma 1}
\begin{equation}\label{L3}
\begin{split}
    \prod_{i=1}^{L_Y}\prod_{j=1}^{L_W}\frac{1-q^{x+Y_i-W_j+j-i}}{1-q^{x+j-i}}
        &\prodp{Y}\frac{1}{1-q^{x+L_W-i+j}}
        \prodp{W}\frac{1}{1-q^{x-L_Y+i-j}}\\
        =&\left.\frac{1}{\cG_{Y,W}(-x,q^{-1})\cG_{W,Y}(x,q)}\right|_{\beta=1}.
    \end{split}
\end{equation}
~\\ \textbf{Proof}

We denote $\cL_n$ as the $n$th term on the left hand side of \eqref{L3} and $\cL_{i,j}$ for the $j$th term of $\cL_i$. $\cL_1$ can be divided into three parts
$$
\cL_1=\prod_{i=1}^{l_Y} \prod_{j=1}^{L_W} \frac{1-q^{x+Y_i-W_j+j-i}}{1-q^{x+j-i}} \prod_{i=l_Y+1}^{L_Y} \prod_{j=1}^{l_W} \frac{1-q^{x-W_j+j-i}}{1-q^{x+j-i}} \prod_{i=1}^{l_Y} \prod_{j=l_W+1}^{L_W} \frac{1-q^{x+Y_i+j-i}}{1-q^{x+j-i}}.
$$
We rewrite $\cL_{1,2}$ as follows
$$
\cL_{1,2}=\prod_{i=l_Y+1}^{L_Y} \prod_{j=1}^{l_W} \prod_{k=1}^{W_j} \frac{1-q^{x+j-i-k} }{1-q^{x+j-i-k+1 }}=\prod_{j=1}^{l_W} \prod_{k=1}^{W_j} \frac{1-q^{x+j-L_Y-k}}{1-q^{x+j-l_Y-k}}.
$$
In the same way, we can get
$$
\cL_{1,3}=\prod_{i=1}^{l_Y} \prod_{k=1}^{Y_i} \frac{1-q^{x-i+L_W+k}}{1-q^{x-i+l_W+k}} .
$$
Notice that the numerators in $\cL_ {1,2}$ and $\cL_ {1,3}$ are equal to the denominators in $\cL_3 $ and $\cL_2$ respectively. Thus, they can both be eliminated. The left hand side of \eqref{L3} becomes
$$
\begin{aligned}
& \text {LHS}=\cL_1 \cL_2 \cL_3=\prod_{i=1}^{l_Y} \prod_{j=1}^{l_W} \frac{1-q^{x+Y_i-W_j+j-i} }{1-q^{x+j-i}}\left(\cL_{1,2} \cL_3\right)\left(\cL_{1,3} \cL_2\right) \\
& ={\prod_{i=1}^{l_Y}}
\prod_{j=1}^{l_W}\frac{1-q^{x+Y_i-W_j+j-i} }{1-q^{x+j-i}} \prod_{i=1}^{l_W } \prod_{j=1}^{W_i} \frac{1}{1-q^{x+i-l_Y-j}} \prod_{i=1}^{l_Y} \prod_{j=1}^{Y_i} \frac{1}{1-q^{x-i+l_W+j}}. \\
&
\end{aligned}
$$
So we need to prove
\be
\label{L3'}
\prod_{i=1}^{l_Y}\prod_{j=1}^{l_W} \frac{1-q^{Y_i-W_j+x+j-i}}{1-q^{x+j-i}}=\prod_{(i, j) \in Y} \frac{1-q^{x-i+l_W+j}}{1-q^{x-W_{i}+j-Y'_j+i-1}}\prod_{(i, j) \in W} \frac{1-q^{x+i-l_Y-j}}{1-q^{x+Y_i+W'_j-i-j+1}}.
\ee

\textbf{The case of $W=\emptyset$.}
$$
\begin{aligned}
& \eqref{L3'}LHS=\prod_{i=1}^{l_Y} \prod_{j=1}^0 \frac{1-q^{Y_i-0+x+j-i}}{1-q^{x+j-i}}=1 ,\\
& \eqref{L3'}RHS = \prod_{(i, j) \in Y} \frac{1-q^{x-i+0+j}}{1-q^{x-0-Y_j^{\prime}+i+j-1}}=\prod_{j=1}^{Y_1} \prod_{i=1}^{Y_j^{\prime}} \frac{1-q^{x-i+j}}{1-q^{x-Y_j^{\prime}+i+j-1}} =
\prod_{j=1}^{Y_1}\prod_{i=1}^{Y_j^{\prime}} \frac{1-q^{x-i+j}}{1-q^{x-i+j}}=1.\\
\end{aligned}
$$
The last equation of RHS is based on the fact that when $j$ is fixed, both $i$ and $Y_j'-i+1$ count from 1 to $Y_j'$.

\textbf{Induction for other cases.}

We assume that Lemma 1 is valid for an arbitrary $W$. We construct partition $C$ which has only one cell different from $W$: $C_m=W_m+1, W_{W_m+1}^{'}=m-1, C_{W_m+1}^{'}=m$, 
where $m$ is the length of W. In particular, $W_m=0$ means $C_m$ starts from a new column, thus we can build any diagram from zero.

We can rewrite \eqref{L3'}
\be\label{L3''}
\prod_{i=1}^{l_Y} \prod_{j=1}^{l_C} \frac{1-q^{Y_i-C_j+x+j-i}}{1-q^{x+j-i}}=\prod_{(i, j) \in Y} \frac{1-q^{x-i+l_C+j}}{1-q^{x-C_i-Y_{j^{\prime}}+i+j-1}} \prod_{(i, j) \in C}
\frac{1-q^{x+i-l_Y-j}}{1-q^{x+Y_i+C_{j^{\prime}}-i-j+1}} .
\ee
We select additional terms in \eqref{L3''} for induction.
$$
\eqref{L3''} LHS=\prod_{i=1}^{l_Y} \prod_{j=1}^{l_W} \frac{1-q^{Y_i-W_j+x+j-i}  }{1-q^{x+j-i}}
\prod_{i=1}^{l_Y}
\frac{1-q^{Y_i-W_m+x+m-i-1}}{1-q^{Y_i-W_m+x+m-i}}.
$$
We use $R_n$ to represent the $n$th term of right hand side of \eqref{L3''}. Considering that $\prod_{(i,j)\in C}f_{i,j}=\prod_{(i,j)\in W}f_{i,j}\times f_{m,W_m+1}$, we obtain
$$
\begin{aligned}
 R_1=&\prod_{(i, j) \in Y} \frac{1-q^{x-i+l_C-j}}{1-q^{x-C_i-Y_j^{\prime}+i+j-1}}=\prod_{(i, j) \in Y} \frac{1-q^{x-i+l_W+j}}{1-q^{x-W_i-Y_j^{\prime}+i+j-1}} \prod_{j=1}^{Y_m} \frac{1-q^{x-Y_j^{\prime}+m+j-W_m-1}}{1-q^{x-Y_j^{\prime}+m+j-W_m-2}}, \\
R_2=&\prod_{(i, j) \in C} \frac{1-q^{x+i-l_Y-j}}{1-q^{x+Y_i+C_j^{\prime}-i-j+1}}=\frac{1-q^{x+m-l_Y-W_m-1}}{1-q^{x+Y_m-W_m}} \prod_{(i, j) \in W} \frac{1-q^{x+i-l_Y-j}}{1-q^{x+Y_i+C_j^{\prime}-i-j+1}}\\=
&
\frac{1-q^{x+m-l_Y-W_m-1}}{1-q^{x+Y_m-W_m}} \prod_{(i, j) \in W} \frac{1-q^{x+i-l_Y-j}}{1-q^{x+Y_i+W_j^{\prime}-i-j+1}} \prod_{i=1}^{m-1} \frac{1-q^{x+Y_i+m-i-W_m-1}}{1-q^{x+Y_i+m-i-W_m}}.
\end{aligned}
$$
Since Lemma 1 is valid for $W$, we just need to prove that
\be\label{L3'''}
\begin{split} 
 &
\prod_{i=1}^{l_Y} \frac{1-q^{x+Y_i-W_m+m-i-1}}{1-q^{x+Y_i-W_m+m-i}} \\
 =&\prod_{j=1}^{Y_m } \frac{1-q^{x-Y_j^{\prime}+m+j-W_m-1}}{1-q^{x-Y_j^{\prime}+m+j-W_m-2}} \times \frac{1-q^{x+m-l_Y-W_m-1}}{1-q^{x+Y_m-W_m}} \prod_{i=1}^{m-1} \frac{1-q^{x+Y_i+m-i-W_m-1}}{1-q^{x+Y_i+m-i-W_m}} .  
\end{split}
\ee
The LHS of \eqref{L3'''} is similar to the third term of RHS, except for the times of product. In the case of $m \geqslant l_Y $, then LHS of \eqref{L3'''} is completely cancelled. Meanwhile, for all $i\geq l_Y $, $Y_i =0$, causing the first term of RHS to disappear. Therefore
$$
\begin{aligned}
    \eqref{L3'''}RHS=&\frac{1-q^{x+m-l_Y-W_m-1}}{1-q^{x-W_m}}\prod_{i=l_Y+1}^{m-1}\frac{1-q^{x+m-i-W_m-1}}{1-q^{x+m-i-W_m}}\\
    =&\frac{1-q^{x+m-l_Y-W_m-1}}{1-q^{x-W_m}}\frac{1-q^{x+m-m+1-W_m-1}}{1-q^{x+m-l_Y-W_m-1}}=1.
\end{aligned}
$$
In the case of $l_Y\geq m$, We can use the same method introduced in Appendix C of \cite{yuanProofA_AGT2023} to complete the recursion. This finishes the proof of Lemma 1.

\bibliographystyle{JHEP}
\bibliography{5D_AGT}

\providecommand{\href}[2]{#2}\begingroup\raggedright\begin{thebibliography}{10}

\bibitem{aldayLiouvilleCorrelationFunctions2010}
L.~F. Alday, D.~Gaiotto and Y.~Tachikawa, \emph{Liouville {{Correlation
  Functions}} from {{Four-dimensional Gauge Theories}}},
  \href{http://dx.doi.org/10.1007/s11005-010-0369-5}{\emph{Lett. Math. Phys.}
  {\bf 91} (2010) 167--197}, [\href{http://arxiv.org/abs/0906.3219}{{\tt
  0906.3219}}].

\bibitem{LeFloch:2020uop}
B.~Le~Floch, \emph{{A slow review of the AGT correspondence}},
  \href{http://dx.doi.org/10.1088/1751-8121/ac5945}{\emph{J. Phys. A} {\bf 55}
  (2022) 353002}, [\href{http://arxiv.org/abs/2006.14025}{{\tt 2006.14025}}].

\bibitem{Awata:2009ur}
H.~Awata and Y.~Yamada, \emph{{Five-dimensional AGT Conjecture and the Deformed
  Virasoro Algebra}},
  \href{http://dx.doi.org/10.1007/JHEP01(2010)125}{\emph{JHEP} {\bf 01} (2010)
  125}, [\href{http://arxiv.org/abs/0910.4431}{{\tt 0910.4431}}].

\bibitem{aganagicGaugeLiouvilleTriality2013}
M.~Aganagic, N.~Haouzi, C.~Kozcaz and S.~Shakirov, \emph{{Gauge/Liouville
  Triality}},  \href{http://arxiv.org/abs/1309.1687}{{\tt 1309.1687}}.

\bibitem{aganagicA_nTriality2014}
M.~Aganagic, N.~Haouzi and S.~Shakirov, \emph{${A}_n$-{{Triality}}},
  \href{http://arxiv.org/abs/1403.3657}{{\tt 1403.3657}}.

\bibitem{Mironov:2010pi}
A.~Mironov, A.~Morozov and S.~Shakirov, \emph{{A direct proof of AGT conjecture
  at beta = 1}}, \href{http://dx.doi.org/10.1007/JHEP02(2011)067}{\emph{JHEP}
  {\bf 02} (2011) 067}, [\href{http://arxiv.org/abs/1012.3137}{{\tt
  1012.3137}}].

\bibitem{albaCombinatorialExpansionConformal2011}
V.~A. Alba, V.~A. Fateev, A.~V. Litvinov and G.~M. Tarnopolsky, \emph{On
  combinatorial expansion of the conformal blocks arising from {{AGT}}
  conjecture}, \href{http://dx.doi.org/10.1007/s11005-011-0503-z}{\emph{Lett.
  Math. Phys.} {\bf 98} (2011) 33--64},
  [\href{http://arxiv.org/abs/1012.1312}{{\tt 1012.1312}}].

\bibitem{Kanno:2013aha}
S.~Kanno, Y.~Matsuo and H.~Zhang, \emph{{Extended Conformal Symmetry and
  Recursion Formulae for Nekrasov Partition Function}},
  \href{http://dx.doi.org/10.1007/JHEP08(2013)028}{\emph{JHEP} {\bf 08} (2013)
  028}, [\href{http://arxiv.org/abs/1306.1523}{{\tt 1306.1523}}].

\bibitem{yuanProofA_AGT2023}
Q.-J. Yuan, S.-P. Hu, Z.-H. Huang and K.~Zhang, \emph{{Proof of $A_{n}$ AGT
  conjecture at $\beta=1$}},  \href{http://arxiv.org/abs/2305.11839}{{\tt
  2305.11839}}.

\bibitem{zhangSelbergIntegralSU2011}
H.~Zhang and Y.~Matsuo, \emph{Selberg integral and {{SU}}({{N}}) {{AGT}}
  conjecture}, \href{http://dx.doi.org/10.1007/JHEP12(2011)106}{\emph{J. High
  Energ. Phys.} {\bf 12} (2011) 106},
  [\href{http://arxiv.org/abs/1110.5255}{{\tt 1110.5255}}].

\bibitem{albionAFLTtypeSelbergIntegrals2021}
S.~P. Albion, E.~M. Rains and S.~O. Warnaar, \emph{{{AFLT-type Selberg}}
  integrals}, \href{http://dx.doi.org/10.1007/s00220-021-04157-0}{\emph{Commun.
  Math. Phys.} {\bf 388} (2021) 735--791},
  [\href{http://arxiv.org/abs/2001.05637}{{\tt 2001.05637}}].

\bibitem{mironovProvingAGTConjecture2012}
A.~Mironov, A.~Morozov, {\relax Sh}.~Shakirov and A.~Smirnov, \emph{Proving
  {{AGT}} conjecture as {{HS}} duality: {{Extension}} to five dimensions},
  \href{http://dx.doi.org/10.1016/j.nuclphysb.2011.09.021}{\emph{Nucl. Phys. B}
  {\bf 855} (2012) 128--151}, [\href{http://arxiv.org/abs/1105.0948}{{\tt
  1105.0948}}].

\bibitem{awataFivedimensionalAGTConjecture2010}
H.~Awata and Y.~Yamada, \emph{Five-dimensional {{AGT Conjecture}} and the
  {{Deformed Virasoro Algebra}}},
  \href{http://dx.doi.org/10.1007/JHEP01(2010)125}{\emph{J. High Energ. Phys.}
  {\bf 2010} (2010) 125}, [\href{http://arxiv.org/abs/0910.4431}{{\tt
  0910.4431}}].

\bibitem{Fukuda:2019ywe}
M.~Fukuda, Y.~Ohkubo and J.~Shiraishi, \emph{{Generalized Macdonald Functions
  on Fock Tensor Spaces and Duality Formula for Changing Preferred Direction}},
  \href{http://dx.doi.org/10.1007/s00220-020-03872-4}{\emph{Commun. Math.
  Phys.} {\bf 380} (2020) 1--70}, [\href{http://arxiv.org/abs/1903.05905}{{\tt
  1903.05905}}].

\bibitem{Zenkevich:2014lca}
Y.~Zenkevich, \emph{{Generalized Macdonald polynomials, spectral duality for
  conformal blocks and AGT correspondence in five dimensions}},
  \href{http://dx.doi.org/10.1007/JHEP05(2015)131}{\emph{JHEP} {\bf 05} (2015)
  131}, [\href{http://arxiv.org/abs/1412.8592}{{\tt 1412.8592}}].

\bibitem{Taki:2014fva}
M.~Taki, \emph{{On AGT-W Conjecture and q-Deformed W-Algebra}},
  \href{http://arxiv.org/abs/1403.7016}{{\tt 1403.7016}}.

\bibitem{Ding:1996mq}
J.~Ding and K.~Iohara, \emph{{Generalization and deformation of Drinfeld
  quantum affine algebras}},
  \href{http://dx.doi.org/10.1023/A:1007341410987}{\emph{Lett. Math. Phys.}
  {\bf 41} (1997) 181--193}, [\href{http://arxiv.org/abs/q-alg/9608002}{{\tt
  q-alg/9608002}}].

\bibitem{miki2007q}
K.~Miki, \emph{A (q, $\gamma$) analog of the $w_{1+\infty}$ algebra},
  {\emph{Journal of Mathematical Physics} {\bf 48} (2007) }.

\bibitem{Ragoucy:1998bv}
E.~Ragoucy and P.~Sorba, \emph{{Yangian realizations from finite W algebras}},
  \href{http://dx.doi.org/10.1007/s002200050034}{\emph{Commun. Math. Phys.}
  {\bf 203} (1999) 551--572}, [\href{http://arxiv.org/abs/hep-th/9803243}{{\tt
  hep-th/9803243}}].

\bibitem{Brundan:2004ca}
J.~Brundan and A.~Kleshchev, \emph{{Shifted Yangians and finite W-algebras}},
  \href{http://arxiv.org/abs/math/0407012}{{\tt math/0407012}}.

\bibitem{Matsuo:2023lky}
Y.~Matsuo, S.~Nawata, G.~Noshita and R.-D. Zhu, \emph{{Quantum toroidal
  algebras and solvable structures in gauge/string theory}},
  \href{http://dx.doi.org/10.1016/j.physrep.2023.12.003}{\emph{Phys. Rept.}
  {\bf 1055} (2024) 1--144}, [\href{http://arxiv.org/abs/2309.07596}{{\tt
  2309.07596}}].

\bibitem{awataFivedimensionalAGTRelation2010}
H.~Awata and Y.~Yamada, \emph{Five-dimensional {{AGT Relation}} and the
  {{Deformed}} beta-ensemble},
  \href{http://dx.doi.org/10.1143/PTP.124.227}{\emph{Prog. Theor. Phys.} {\bf
  124} (2010) 227--262}, [\href{http://arxiv.org/abs/1004.5122}{{\tt
  1004.5122}}].

\bibitem{awataRefinedBPSState2009}
H.~Awata and H.~Kanno, \emph{Refined {{BPS}} state counting from {{Nekrasov}}'s
  formula and {{Macdonald}} functions},
  \href{http://dx.doi.org/10.1142/S0217751X09043006}{\emph{Int. J. Mod. Phys.
  A} {\bf 24} (2009) 2253--2306}, [\href{http://arxiv.org/abs/0805.0191}{{\tt
  0805.0191}}].

\bibitem{selberg1944bemerkninger}
A.~Selberg, \emph{Bemerkninger om et multiplet integral}, {\emph{Norsk. Mat.
  Tidsskr.} {\bf 24} (1944) 71--78}.

\bibitem{kadell1988proof}
K.~W. Kadell, \emph{A proof of some q-analogues of selberg’s integral for
  k=1}, {\emph{SIAM journal on mathematical analysis} {\bf 19} (1988)
  944--968}.

\bibitem{stanley1989some}
R.~P. Stanley, \emph{Some combinatorial properties of jack symmetric
  functions}, {\emph{Advances in Mathematics} {\bf 77} (1989) 76--115}.

\bibitem{albionEllipticMathrm_n2023}
S.~P. Albion, E.~M. Rains and S.~O. Warnaar, \emph{Elliptic ${A}_n$ {{Selberg}}
  integrals},  \href{http://arxiv.org/abs/2306.02442}{{\tt 2306.02442}}.

\bibitem{warnaarSelbergIntegralLie2009}
S.~O. Warnaar, \emph{A {{Selberg}} integral for the {{Lie}} algebra ${A}_n$},
  \href{http://dx.doi.org/10.1007/s11511-009-0043-x}{\emph{Acta Math.} {\bf
  203} (2009) 269--304}, [\href{http://arxiv.org/abs/0708.1193}{{\tt
  0708.1193}}].

\bibitem{askeyBasicHypergeometricExtensions1980}
R.~Askey, \emph{Some {{Basic Hypergeometric Extensions}} of {{Integrals}} of
  {{Selberg}} and {{Andrews}}},
  \href{http://dx.doi.org/10.1137/0511084}{\emph{SIAM J. Math. Anal.} {\bf 11}
  (1980) 938--951}.

\bibitem{forresterImportanceSelbergIntegral2008}
P.~J. Forrester and S.~O. Warnaar, \emph{The importance of the {{Selberg}}
  integral}, \href{http://dx.doi.org/10.1090/S0273-0979-08-01221-4}{\emph{Bull.
  Amer. Math. Soc.} {\bf 45} (2008) 489--489}.

\bibitem{macdonald1995}
I.~G. Macdonald, \emph{Symmetric Functions and Hall Polynomials}.
\newblock Oxford University Press, New York, second~ed., 1995.

\bibitem{comanTodaConformalBlocks2020}
I.~Coman, E.~Pomoni and J.~Teschner, \emph{Toda {{Conformal Blocks}}, {{Quantum
  Groups}}, and {{Flat Connections}}},
  \href{http://dx.doi.org/10.1007/s00220-019-03617-y}{\emph{Commun. Math.
  Phys.} {\bf 375} (2020) 1117--1158}.

\bibitem{fateevCorrelationFunctionsConformal2007}
V.~A. Fateev and A.~V. Litvinov, \emph{Correlation functions in conformal
  {{Toda}} field theory {{I}}},
  \href{http://dx.doi.org/10.1088/1126-6708/2007/11/002}{\emph{J. High Energy
  Phys.} {\bf 2007} (2007) 002--002},
  [\href{http://arxiv.org/abs/0709.3806}{{\tt 0709.3806}}].

\bibitem{fateevCorrelationFunctionsConformal2009}
V.~A. Fateev and A.~V. Litvinov, \emph{Correlation functions in conformal
  {{Toda}} field theory {{II}}},
  \href{http://dx.doi.org/10.1088/1126-6708/2009/01/033}{\emph{J. High Energy
  Phys.} {\bf 2009} (2009) 033--033},
  [\href{http://arxiv.org/abs/0810.3020}{{\tt 0810.3020}}].

\bibitem{dotsenkoConformalAlgebraMultipoint1984}
V.~S. Dotsenko and V.~A. Fateev, \emph{Conformal algebra and multipoint
  correlation functions in 2d statistical models}, {\emph{Nuclear Physics B}
  {\bf 240} (1984) 312--348}.

\bibitem{itoyamaMethodGeneratingQExpansion2010}
H.~Itoyama and T.~Oota, \emph{Method of {{Generating}} q-{{Expansion
  Coefficients}} for {{Conformal Block}} and {{N}}=2 {{Nekrasov Function}} by
  beta-{{Deformed Matrix Model}}},
  \href{http://dx.doi.org/10.1016/j.nuclphysb.2010.05.002}{\emph{Nucl. Phys. B}
  {\bf 838} (2010) 298--330}, [\href{http://arxiv.org/abs/1003.2929}{{\tt
  1003.2929}}].

\bibitem{Mironov:2010zs}
A.~Mironov, A.~Morozov and S.~Shakirov, \emph{{Conformal blocks as
  Dotsenko-Fateev Integral Discriminants}},
  \href{http://dx.doi.org/10.1142/S0217751X10049141}{\emph{Int. J. Mod. Phys.
  A} {\bf 25} (2010) 3173--3207}, [\href{http://arxiv.org/abs/1001.0563}{{\tt
  1001.0563}}].

\end{thebibliography}\endgroup

\end{document}